\documentclass[aps,prb,groupedaddress,showpacs,twocolumn]{revtex4}
\usepackage{graphicx}
\usepackage{amsmath}
\usepackage{amssymb}
\usepackage{bbm}
\usepackage{xspace}
\usepackage{color}
\usepackage{hyperref}
\definecolor{orange}{RGB}{252,77,6}
\definecolor{brown}{RGB}{200,127,50}
\definecolor{green1}{RGB}{00,100,00}
\definecolor{green2}{RGB}{00,150,00}
\definecolor{green3}{RGB}{00,200,00}
\definecolor{green4}{RGB}{00,250,00}

\newcommand{\LIMOO}{Li$_{0.9}$Mo$_{6}$O$_{17}$}

\newcommand{\fig}[1]{Fig.\thinspace{}\ref{#1}}
\newcommand{\eq}[1]{eq.\thinspace{}(\ref{#1})}

\newcommand{\se}{Sec.\@\xspace}

\newcommand{\app}{Appendix\@\xspace}

\newcommand{\etal}[0]{\textit{et al.}}
\newcommand{\tcite}[1]{Ref.~\onlinecite{#1}}
\newcommand{\tcites}[1]{Refs.~\onlinecite{#1}}

\def\bra#1{\mathinner{\langle{#1}|}}
\def\ket#1{\mathinner{|{#1}\rangle}}

\newcommand{\tab}[1]{Tab.\thinspace{}\ref{#1}}

\begin{document}

\title{Effective model for electronic properties of the quasi one-dimensional purple bronze \LIMOO{} based on {\em ab-initio} calculations}

\author{Martin Nuss}
\email[]{martin.nuss@student.tugraz.at}
\affiliation{Institute of Theoretical and Computational Physics, Graz University of Technology, 8010 Graz, Austria}
\author{Markus Aichhorn}
\affiliation{Institute of Theoretical and Computational Physics, Graz University of Technology, 8010 Graz, Austria}

\date{\today}

\begin{abstract}
We investigate the electronic structure of the strongly anisotropic, quasi low dimensional purple bronze \LIMOO. Building on all-electron {\em ab-initio} band structure calculations we obtain an effective model in terms of four maximally localized Wannier orbitals, which turn out to be far from atomic like. We find {\em two half-filled} orbitals arranged in chains running along one crystallographic direction, and {\em two full} orbitals in perpendicular directions, respectively. The possibility to reduce this model to only two orbitals forming two chains per unit cell with inter-chain coupling is discussed. Transport properties of these models show high anisotropy, reproducing trends of the experimentally determined values for the dc conductivity. We also consider basic effects of electron-electron interactions using the (extended) Variational Cluster Approach and Dynamical Mean Field Theory. We find good agreement with experimental photo emission data upon adding moderate on-site interaction of the order of 
the 
band width to the {\em ab-initio} derived tight-binding Hamiltonian. The obtained models provide a profound basis for further investigations on low-energy Luttinger liquid properties or to study electronic correlations within computational many-body theory. 
\end{abstract}

%71.27+a strongly correlated systems
%71.10.-w theories and models of condensed matter
%71.15.-m condensed matter calculation methods
%72.15.Qm Kondo Effect
%71.55.Ak local magnetic moments in metals
%73.63.Kv       Quantum dots
%73.23.-b       Electronic transport in mesoscopic systems
%72.10.Fk Scattering by point defects, dislocations, surfaces, and other imperfections (including Kondo effect)
%71.15.-m condensed matter calculation methods
%72.15.Qm Kondo Effect
%71.27+a strongly correlated systems
%71.10.-w theories and models of condensed matter
% Insulator-metal transitions, 71.30.+h
%Low-dimensional structures devices, 85.35.Be
%crystalline solids, 71.20.-b electrical properties, 73.63.-b
%71.20.-b Electron density of states and band structure of
%crystalline solids
%72.15.-v Electronic conduction in metals and alloys

\maketitle

\section{Introduction}\label{sec:introduction}
% GENERAL MOTIVATION: ELECTRONIC PROPERTIES OF LOW DIMENSIONAL
% MATERIALS
The electronic structure of highly anisotropic materials shows a
plethora of interesting effects. Quantum many-body
dynamics in quasi low dimensional systems becomes important and dominant
in many regions of their rich phase diagram. 
This often implies unconventional ground states, such as
non-Fermi-liquid or Luttinger-liquid states. One prominent example for this class of materials is the lithium molybdenum
purple bronze \LIMOO,~\cite{McCarroll1984282} a molybdenum oxide
bronze with quasi one-dimensional properties.~\cite{doi:10.1021/cr00083a002}

% LIMOO ELECTRONICS IS LOW DIMENSIONAL 1: STRUCTURE AND TRANSPORT
Experimental structure analysis using X-rays~\cite{Onoda1987163}
as well as neutrons~\cite{PhysRevB.84.014108} determined a monoclinic
crystal structure. The conduction electrons are mostly located on two
molybdenum octahedral sites which are arranged in double zig-zag chains
along the $b$ axis. This leads to a very high anisotropy of the
material, which has been studied by several techniques using
resistivity measurements,~\cite{Greenblatt1984671,PhysRevB.76.233105,
  PhysRevLett.108.187003,PhysRevLett.102.206602,PhysRevB.77.193106,0022-3719-19-30-014}
conductivity under pressure,~\cite{EscribeFilippini1989427} magneto
resistance,~\cite{PhysRevLett.102.206602,0295-5075-89-6-67010} thermal
expansion,~\cite{PhysRevLett.98.266405} optical
conductivity,~\cite{PhysRevB.69.085120,PhysRevB.38.5821} the Nernst
effect,~\cite{PhysRevLett.108.056604} thermal
conductivity,\cite{NatComm.2.396} thermopower~\cite{Boujida1988465}
and muon spectroscopy.~\cite{Chakhalian20051333}

% LIMOO ELECTRONICS IS LOW DIMENSIONAL 2: ELECTRON DYNAMICS
The electronic properties have been addressed using angle-resolved photo emission spectroscopy
(ARPES)~\cite{PhysRevLett.82.2540,PhysRevLett.96.196403,PhysRevLett.103.136401,PhysRevB.68.195117,PhysRevB.70.153103,Wang20081490,0953-8984-25-1-014007,Gweon2002584,PhysRevLett.85.3985}
and scanning tunneling microscopy
(STM),~\cite{PhysRevLett.95.186402,0953-8984-25-1-014008} which argued
for one-dimensional Luttinger liquid physics.~\cite{doi:10.1021/cr030647c,PhysRevLett.96.196405,PhysRevB.76.155402,PhysRevB.46.15753,PhysRevB.47.6740,PhysRevB.75.195123,PhysRevLett.85.3985} Other studies disputed this claim.~\cite{PhysRevLett.83.1235} The evolution and the current status of work in that direction is summed up in a recent review article.~\cite{0953-8984-25-1-014007} A
temperature dependent dimensional
crossover~\cite{PhysRevLett.87.276405,PhysRevLett.97.136401,PhysRevLett.109.126404}
which induces coherence for the perpendicular electron motion, has been
studied using neutron diffraction.~\cite{PhysRevB.84.014108}

% LIMOO ELECTRONICS: ``PHASES``
Apart from intriguing physical effects of effective low dimensionality
the material shows superconductivity below
$1.9$\,K~\cite{Schlenker1985511,doi:10.1021/cr00083a002,Matsuda1986243,Ekino198741,Mercure2012}
and a metal-insulator transition is observed at around
$24$\,K.~\cite{Schlenker1985511, Greenblatt1984671, 0022-3719-20-9-007,
  PhysRevB.69.085120} No evidence for a Peierls instability has been
reported,~\cite{Gweon2001481} and a possible charge density wave (CDW)
phase is still under debate.~\cite{doi:10.1142/S0217979293003589,
  PhysRevB.86.075147,PhysRevB.85.235128,PhysRevLett.95.186402} Recent
studies have argued for a compensated metal.~\cite{PhysRevB.86.195143} All data
are summed up in a conjectured electronic phase diagram as presented
in \tcite{PhysRevB.85.235128}. 

\begin{figure*}
\includegraphics[width=0.80\textwidth]{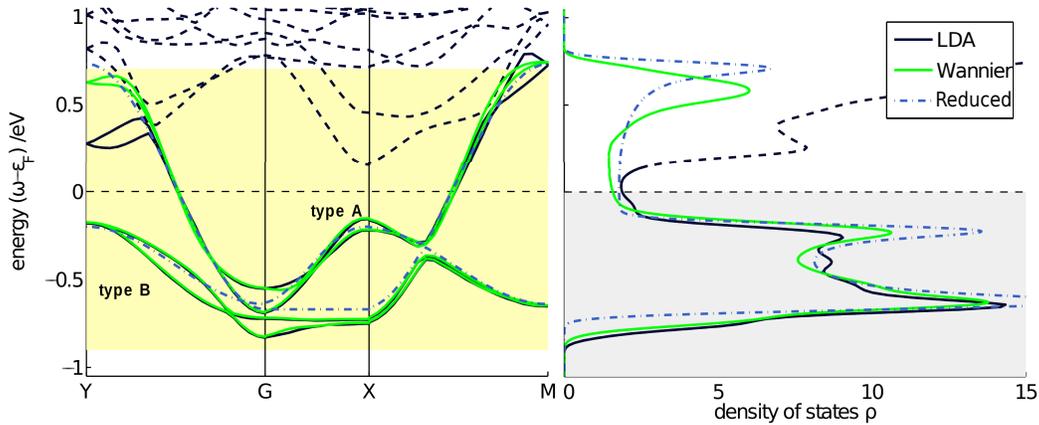}
\caption{(Color online) Left: LDA band structure (solid black) in the
  vicinity of the Fermi energy $\epsilon_F$ plotted along a
  two-dimensional path in the reciprocal $b$-$c$ plane. The four bands
  of the full Wannier projected Hamiltonian are shown on top 
  (solid green) next to the data for the reduced model
  (dash-dotted blue). Right: DOS of the LDA
  calculation (black), the full Wannier model (solid green) and the reduced
  model (dash-dotted blue).}  
\label{fig:bsDos}
\end{figure*}

% THEORY: LIMOO ELECTRONICS: SURVEY OF THEORY
Theoretical {\em ab-initio} studies of the electronic structure using
a tight-binding method,~\cite{doi:10.1021/ja00210a006} as well as a
linearized muffin-tin orbital
(LMTO)~\cite{PhysRevB.74.045117,PhysRevB.85.235108} 
calculation in the
local density approximation (LDA) have been
conducted. These approaches were successful in providing a broad picture
of the ''high energy'' physics of \LIMOO, accounting for the high
anisotropy. Although experiments testified a wealth of remarkable
low-energy properties and different quantum ground states, more
detailed 
theoretical investigations, including interactions and low
dimensionality, emerged in recent years only. 

% THEORY:  LIMOO ELECTRONICS: LUTTINGER
Chudzinski \etal~\cite{PhysRevB.86.075147} investigated the quasi
one-dimensionality and have been able to extract an effective {\em low-energy} theory within the Tomonaga-Luttinger-liquid framework. Their
approach is based on an atomic orbital tight-binding model with
parameters such that it matches an LDA LMTO band-structure
calculation. Motivated by the crystal structure of \LIMOO{} the model
was set up with four molybdenum $d$ orbitals in a zig-zag ladder
arrangement including on-site as well as non-local electronic
interactions. It was found that within this model Luttinger-liquid
low-energy parameters can be obtained, which are consistent with
experimental findings.  

% THEORY:  LIMOO ELECTRONICS: UNIVERSAL MODEL
Another recent work~\cite{PhysRevB.85.235128} proposes a
two-dimensional model from {\em Slater-Koster}~\cite{PhysRev.94.1498} 
atomic orbitals also including non-local electronic
interactions. Again an Ansatz with four Mo orbitals in zig-zag ladder
arrangement was applied. The authors argue, based on electron counting,
that there are two electrons to be shared among the four equivalent Mo
atoms, leading to quarter-filled orbitals. The bandwidth obtained with
this Ansatz for the two bands crossing the Fermi level is in rough
agreement with Density Functional Theory (DFT) 
calculations. Details of the band structure such as curvatures,
however, and also the bands just below the Fermi level which are of
similar Mo $d$ character, cannot be reproduced by this {\em Slater-Koster}
model.  

% THEORY: WHERE WE COME IN: MODEL
The main purpose of this work is to establish an unbiased, {\em
  general purpose tight-binding model} for the electronic properties of \LIMOO{} based
on {\em ab-initio} calculations. Such a model is intended to serve as a
basis to study the role of electronic correlations by adding
interactions, be it in a computational many-body theory or in a
one-dimensional renormalization group (RG) framework. In contrast to 
previous work~\cite{PhysRevB.85.235128} we propose a model based on
maximally-localized Wannier
orbitals~\cite{RevModPhys.84.1419,Marzari2003} instead of linear
combination of atomic orbitals. Four molecular-like orbitals are
obtained in a fully {\em ab-initio} approach from an all-electron DFT
calculation. Our results unambiguously show that, using a set of
four Wannier orbitals in the unit cell, the model consists of {\em two half-filled} as well as {\em two filled} orbitals. As we will show
below, the DFT band structure is perfectly reproduced in this basis
set of Wannier functions. 

% THEORY: CORRELATIONS GENERAL
This model describes the momentum resolved electronic
structure as observed in ARPES~\cite{PhysRevLett.103.136401}
experiments and reproduces highly anisotropic transport
characteristics.~\cite{Greenblatt1984671,0295-5075-89-6-67010,PhysRevB.69.085120,PhysRevB.76.233105,PhysRevLett.108.187003,PhysRevLett.102.206602,NatComm.2.396} Furthermore 
we discuss an even simpler two-orbital effective model which can be
derived from the four-orbital model. 

% THEORY: CORRELATIONS GENERAL
In the second part of the paper we conduct a first (qualitative) study
of effects of 
interactions on the electron dynamics within this effective Wannier
model. Even more so due to the low 
dimensionality the interacting model is in general difficult to
approach. By applying RG as well as density matrix renormalization group
(DMRG)~\cite{Schollwock201196} in certain limits (chains, ladders) their essential physics can
be understood.~\cite{PhysRevB.85.235128} To solve the full low 
dimensional interacting Hubbard-type model, general
frameworks as for example (cluster) dynamical mean field theory ((C)DMFT)-like
approaches~\cite{RevModPhys.77.1027} have been applied, where the self-energy of the system is
restricted to a finite length scale. 

\begin{figure*}
\includegraphics[width=0.80\textwidth]{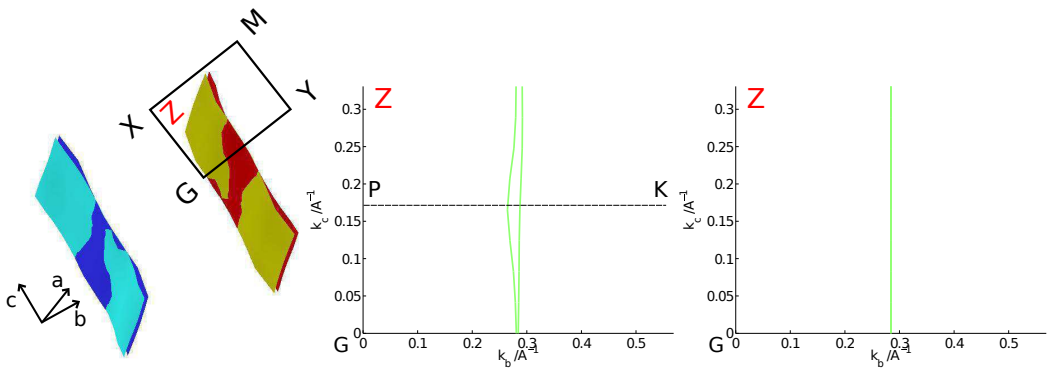}
\caption{(Color online) Calculated Fermi surface. Left: The {\em ab-initio} result
  (image created using XCrySDen~\cite{Kokalj2003155}). Center:
  In-plane projection of the result for the four-orbital Wannier 
  model. Right: In-plane projection of the result for the reduced
  model, where type (A) orbitals are strictly one-dimensional and the
  two Fermi sheets are degenerate.}
\label{fig:FS}
\end{figure*}

% THEORY: CORRELATIONS WHAT WE DO
Realistic modeling is a relatively new and rapidly developing field.~\cite{0953-8984-20-29-293201,JPSJ.79.112001,doi:10.1142/S0217979201006495} In this work we study (simple) electron-electron interactions in the
effective model using complementary numerical techniques. First, we
use cluster perturbation theory
(CPT)~\cite{gros_cluster_1993, senechal_spectral_2000} as well as the
(extended)~\cite{PhysRevB.70.235107,PhysRevB.72.155110} Variational
Cluster Approach~\cite{potthoff_variational_2003} ((e)VCA) in the
spirit of LDA+VCA.~\cite{PhysRevB.75.140406,PhysRevB.80.115129} The
choice of these methods is motivated by the expected reduced
effective dimensionality of the material which renders the non-local
character of the VCA self-energy an interesting perspective.
Second, we apply the well-established LDA+DMFT~\cite{0953-8984-9-35-010,0953-8984-11-4-011,PhysRevB.68.144425} 
approach, which neglects non-local correlations, but on the other hand
performs superior in describing the quasi-particle features at low
energy as compared to VCA. 
For all applied methods we find that a moderate value of
on-site interactions strength is capable of describing the electron
dynamics best and in good agreement with ARPES experiments. We discuss
the influence of a hybridization mechanism of the 
two bands right at the Fermi energy with the two bands slightly below,
not accounted for in previous work. 

% OUTLINE
This paper is organized as follows: in \se~\ref{sec:model} we report
accurate all electron DFT data from which we obtain a model in terms
of maximally-localized Wannier functions. A further simplified
model for \LIMOO{} with reduced number of hopping parameters is discussed in
\se~\ref{ssec:minimalmodel}. We 
present results for the anisotropic conductivity in
\se~\ref{sec:anisotropicconductivity} and compare to transport
measurements. The electron dynamics of the interacting effective model
are presented and compared to ARPES experiments in
\se~\ref{sec:correlatedelectronicstructure} before concluding in \se~\ref{sec:conclusions}. 

\section{From crystal structure to an effective electronic model}\label{sec:model}

\subsection{{\em Ab-initio} electronic structure}\label{ssec:abinitio}

\begin{figure}
\centering
\includegraphics[width=0.4\textwidth]{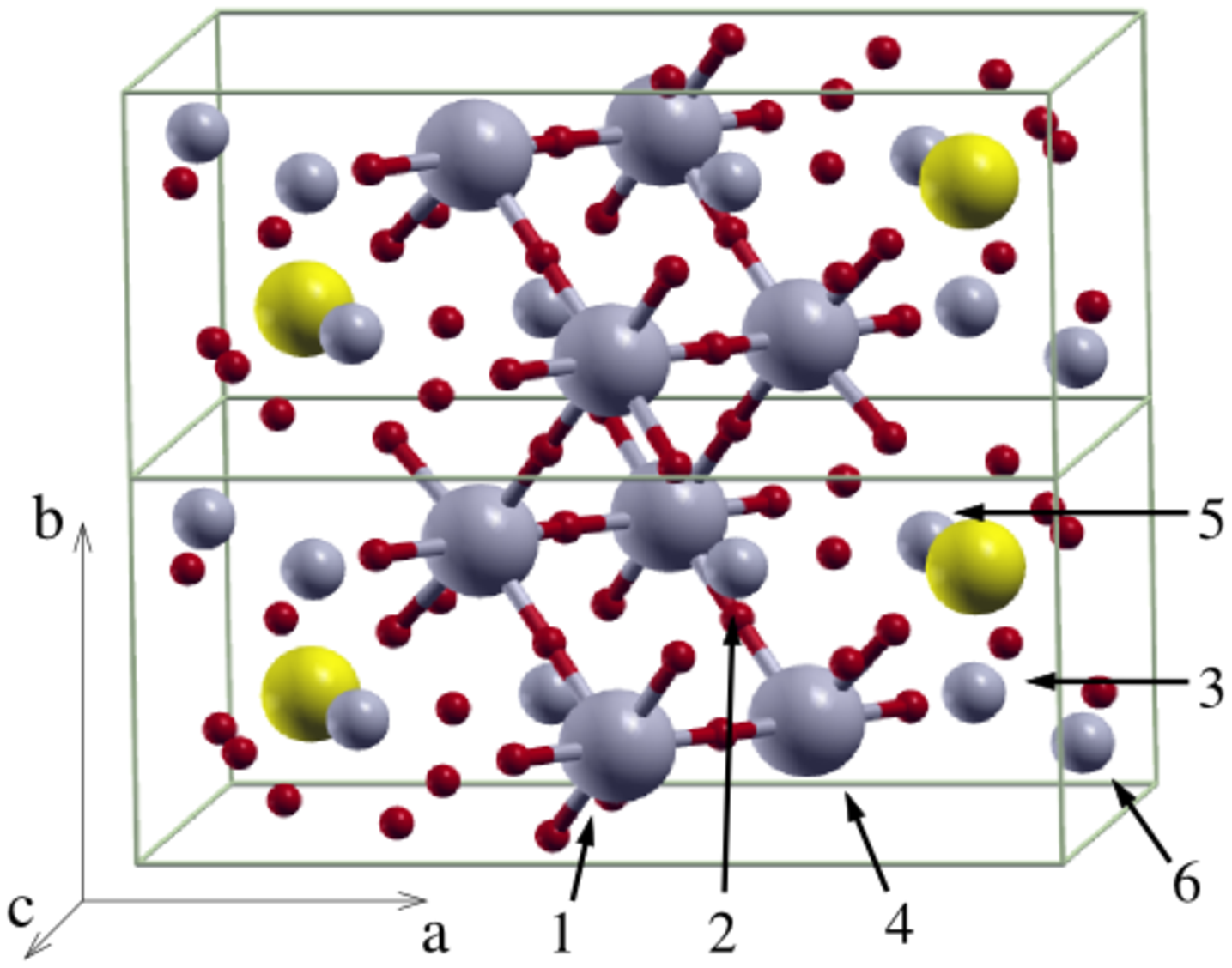}\\
\includegraphics[width=0.35\textwidth]{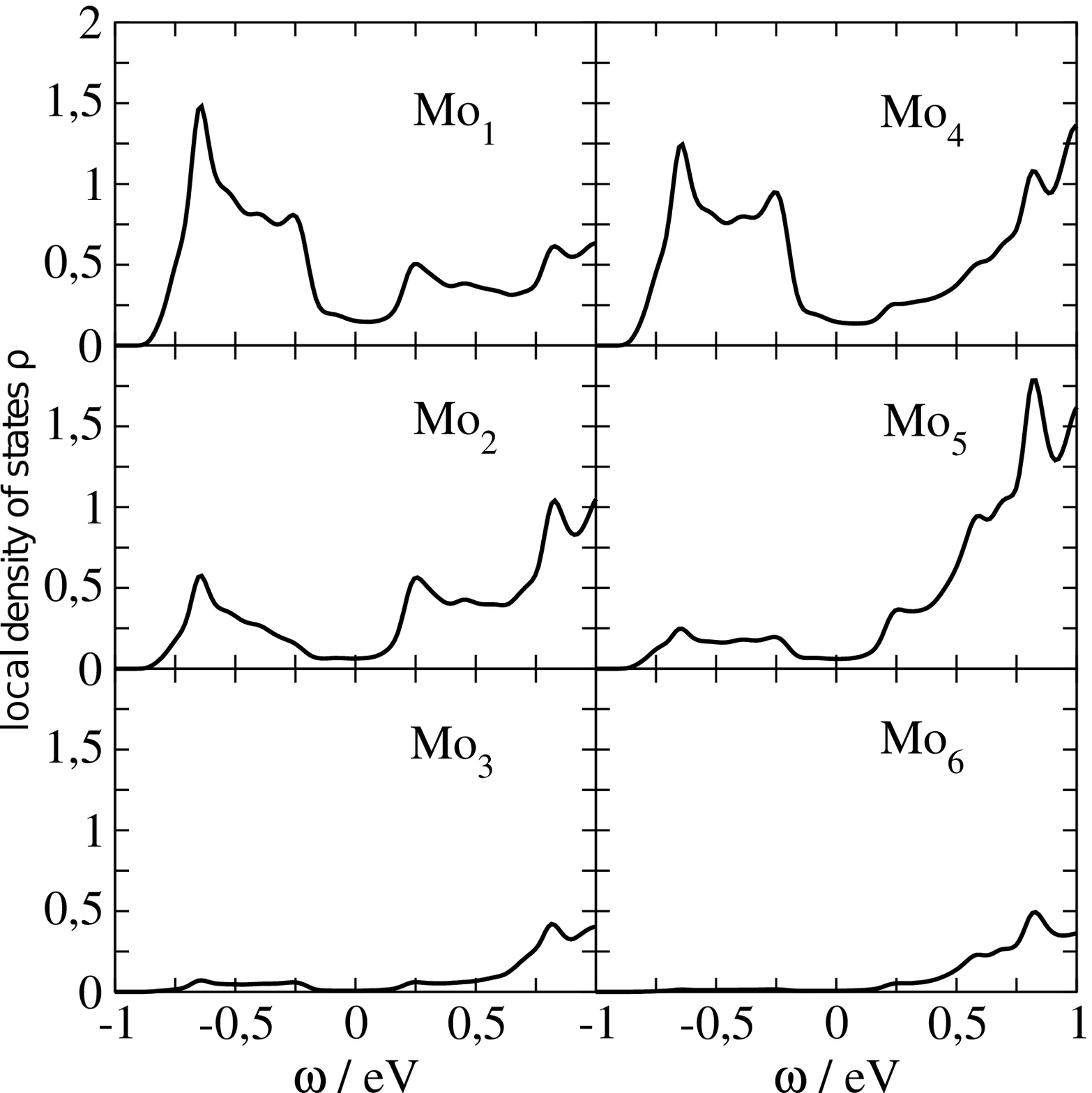}
\caption{ Top plot: Crystal structure. In $a$ and $c$ direction
  one unit cell, and in $b$ direction 2 unit cells are shown. Big gray
  balls are Mo$_1$/Mo$_1^\prime$ and Mo$_4$/Mo$_4^\prime$ atoms, showing the zig-zag chain structure
  along $b$. Small gray: other Mo sites. The numbers next to the black
  arrows denote the atom number. 
  Small red balls are oxygen,
  and the yellow balls Li atoms. This image has been created using
  XCrySDen.~\cite{Kokalj2003155} Bottom plot: Partial DOS
  of the six in-equivalent Mo atoms in the unit cell. Top row: atoms
  Mo$_1$ and Mo$_4$, forming the zig-zag 
  chains. Middle row: Mo$_2$ and Mo$_5$. Bottom row: Mo$_3$ and Mo$_6$.}  
\label{fig:partialDOS_structure}
\end{figure}

\begin{figure}
\includegraphics[width=0.45\textwidth]{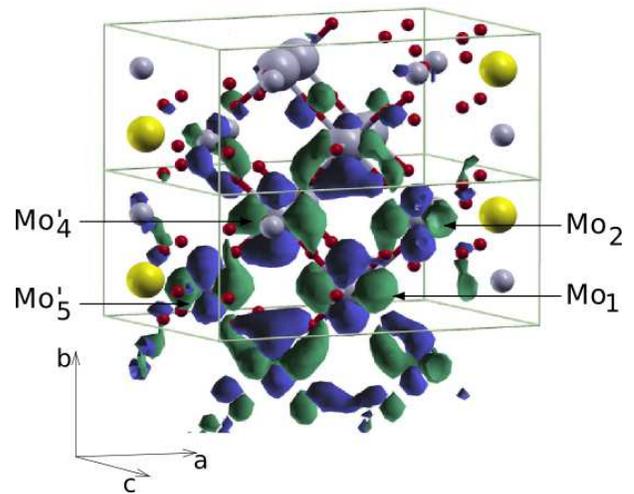}
\caption{Visualization of one type (A) Wannier orbital. Similar as in
  \fig{fig:partialDOS_structure}, two unit cells in
  $b$ direction are shown, the Wannier function is centered in the
  lower unit cell. The center of the Wannier function is located in between
  atoms Mo$_1$ and Mo$_4^\prime$. Arrows mark the atoms with
  significant contribution to the Wannier function: in-chain atoms
  Mo$_1$ and Mo$_4^\prime$, and adjacent atoms Mo$_2$ and
  Mo$_5^\prime$. Color coding as in the top panel of  
  \fig{fig:partialDOS_structure}. For the Wannier functions, blue and green lobes denote
  positive and negative phase, resp. The image has been created using XCrySDen.~\cite{Kokalj2003155}}
\label{fig:latticeWannier}
\end{figure}

% AB-INITIO METHOD + CRYSTAL STRUCTURE
We obtain the electronic structure for ideal Li$_{1}$Mo$_{6}$O$_{17}$
from a non-spin-polarized, full-potential linearized augmented plane wave (FP-LAPW)~\cite{PhysRev.51.846,PhysRevB.12.3060,PhysRevB.43.6388,Sjostedt200015}
DFT~\cite{PhysRev.136.B864,PhysRev.140.A1133} calculation as
implemented in the WIEN2k package.~\cite{blaha_wien2k_2001} The
unit-cell parameters and crystal structure are taken from X-ray 
data,\cite{Onoda1987163} which have been recently confirmed by
neutron diffraction experiments.~\cite{PhysRevB.84.014108} The space
group is  monoclinic (prismatic) $P2_1/m$ with lattice parameters $a
=12.762(2)$\,\AA, $b = 5.523(1)$\,\AA, $c = 9.499(1)$\,\AA, $\beta =
90.61(1)^\circ$ and $Z = 2$, leading to a $48$-atom unit cell
[Li$_{1}$Mo$_{6}$O$_{17}$]$_2$.~\cite{footnote1} 

% AB-INITIO NUMERICAL PARAMETERS
All results presented in this work are calculated with the
exchange-correlation potential treated in the 
LDA.~\cite{PhysRevLett.45.566} We checked that the
generalized-gradient approximation (GGA -
PBE~\cite{PhysRevB.54.16533}) gives indistinguishable results for the
band structures. 
Our results are converged in terms of the size of
the FP-LAPW basis set, which is determined by the $RK_{\text{max}}$
parameter in WIEN2k. By performing calculations for different
$RK_{\text{max}}$ we found that $RK_{\text{max}}=7.0$ and
$RK_{\text{max}}=6.0$ gave the same results, with band energies within
$10^{-3}\,$eV, only at $RK_{\text{max}}=5.0$ deviations become
visible. Therefore, also due to the computational complexity of the
problem, all results presented here are obtained with a
$RK_{\text{max}}=6.0$ basis set. 

% OBTAINED ELECTRONIC STRUCTURE GENERAL
The obtained electronic structure
$\mathbf{\epsilon}_{\text{KS}}(\mathbf{k})$ is visualized along the
standard path of the $b$-$c$ plane in reciprocal space (Y-G-X-M) (see also \fig{fig:FS}) in
\fig{fig:bsDos} (left). 
In order to compare to ARPES experiment, we modeled lithium-vacant
\LIMOO{} by a rigid band shift of $0.03\,$eV of the LDA
bands.~\cite{PhysRevB.74.045117}  
We find a combined bandwidth of the four bands
in the vicinity of the Fermi energy $\epsilon_F$ of $W\approx1.82$\,eV and two Fermi
velocities of $v_{F,1}\approx 0.99\cdot10^5\,\frac{\text{m}}{\text{s}}$ and $v_{F,2}\approx 0.93\cdot10^5\,\frac{\text{m}}{\text{s}}$, roughly one order of magnitude lower than in free electron metals.~\cite{ashcroft_1976} The corresponding electronic
density of states (DOS) is shown in \fig{fig:bsDos} (right). The LDA DOS is
obtained by Gaussian integration ($\sigma\approx0.02$\,eV) using the
tetrahedron method on a grid of $216$ $k$-points in
the irreducible Brillouin zone (BZ). 

% OBTAINED ELECTRONIC STRUCTURE GENERAL IN CONTEXT WITH PREV. WORK
By and large the electronic structure compares well to previous
data reported in early works of Whangbo
\etal~\cite{doi:10.1021/ja00210a006} from an empirical tight-binding
method, and also to more recent LMTO calculations within the atomic sphere
approximation (ASA) from Popovic \etal~\cite{PhysRevB.74.045117} But
note that in particular the band
crossings/hybridizations on the X-M line as well as the lowest empty
bands at $\sim1$\,eV above $\epsilon_F$ are
apparently different. Since we checked accurately the convergence of
the all-electron FP-LAPW calculations the difference is most likely to come
from the approximations introduced in LMTO-ASA and tight-binding
calculations.

% FERMI SURFACE
The one-dimensionality of the material
becomes manifest in the Fermi surface which is shown in \fig{fig:FS}
(left). Arising from two bands crossing the Fermi energy, it consists
of two sheets warping in $c$ direction, cutting the $b$ axis and
being roughly constant in $a$ direction. In experiments~\cite{Allen2013} the maximum splitting of the Fermi surface $(<10^{-3}\AA^{-1})$ is observed along the $\overline{\text{PK}}$ line. Our LDA calculations yield the maximum splitting along the very same line, see \fig{fig:FS}, but the magnitude is much larger ($\approx0.021\AA^{-1}$). In previous LDA calculations~\cite{PhysRevB.74.045117} an even larger splitting of $\approx0.045\AA^{-1}$ was found. This discrepancy of the theoretical results with experiment is likely due to the improper treatment of strong non-local electronic correlations in the LDA.

\subsection{Realistic effective model}\label{ssec:abinitiomodel}

% OBATAINED ELECTRONIC STRUCTURE - IDEA FOR MODEL
To construct an effective model we have to identify the
  origin (orbital character and atom) of those electronic states which are most important for the physical properties i.e. those close to the Fermi energy $\epsilon_F$. We plot in the bottom panel of \fig{fig:partialDOS_structure} the partial DOS for the six in-equivalent Mo atoms in the 
  unit cell. One can nicely see that Mo$_1$ and Mo$_4$ contribute most
  to the DOS at $\epsilon_F$ (for nomenclature see Figure 2 in Onoda {\em et
    al.}~\cite{Onoda1987163}). In
  the top panel of \fig{fig:partialDOS_structure} we show the
  crystal structure, with emphasis on those Mo$_1$ and Mo$_4$ (including
  the equivalent Mo$_1^\prime$ and Mo$_4^\prime$) atoms. It is
  evident that these atoms form the two adjacent zig-zag chains
  running along the $b$-axis, giving rise to the two
  quasi-one-dimensional bands crossing $\epsilon_F$. The atoms Mo$_2$ and Mo$_5$
  are sitting next to the chains, and thus have some smaller
  contributions. The other two Mo atoms are far away from the chains, and
  thus contribute hardly anything to the weight around $\epsilon_F$.
  This analysis of the orbital character shows clearly that the bands
  around $\epsilon_F$ originate mainly from only four atoms (Mo$_1$, Mo$_4$, and
  Mo$_1^\prime$, Mo$_4^\prime$, respectively) in the unit cell. 

To construct an effective model, we 
  take the electronic wave function data $\mathbf{\phi}_{\text{KS}}$ in an
  energy window of $[-0.9, 0.7]$\,eV, that comprises the 4 relevant
  bands as shown in \fig{fig:bsDos}. The lower bound of the energy
  window for projection is straightforward to choose because the gap
  between the four considered molybdenum $d$ bands and the next lower
  bands is larger than $1.5$\,eV. The upper bound is more involved since bands with
  different character penetrate the energy window from above, and are
  entangled with the two bands crossing the Fermi energy. In order to
  get a good description of the bands, we had to use the disentanglement
  procedure of Wannier90 with a frozen energy window of $[-0.9, 0.0]$\,eV. 

  We project this data onto four
  maximally-localized Wannier orbitals~\cite{RevModPhys.84.1419}
  $\omega_\alpha$ using Wannier90~\cite{Mostofi2008685} and the
  Wien2Wannier~\cite{Kunes2010} interface. As initial seed, we chose
  one $d_{xy}$ orbital on each of the Mo$_1$, Mo$_4$, Mo$_1^\prime$,
  and Mo$_4^\prime$ atoms.

% WANNIER MODEL OUTCOME
Although starting from a seed with atomic orbitals, the calculated
  Wannier functions, however, have quite 
  different character. They can be divided into two kinds. Type
  (A), which is oriented along chains in $b$ direction, and type (B) which is
  in some sense orthogonal in real space, mediating between the chains
  in $b$ direction. 
  The orbitals contributing to the states around $\epsilon_F$ are of type
  (A), and one of these orbitals is shown in \fig{fig:latticeWannier}. One
  can clearly see the $d_{xy}$ orbital character, forming the zig-zag
  chains, around atoms Mo$_1$
  and Mo$_4^\prime$, where most of the orbital weight is located. In
  consistency with the partial DOS, \fig{fig:partialDOS_structure}, some
  contribution also comes from atoms Mo$_2$ and Mo$_5^\prime$, since
  they are adjacent to the chains, as shown in \fig{fig:latticeWannier}.

The splitting into two types of orbitals can
  be understood from the band structure. Only two bands cross $\epsilon_F$,
  which results in two equivalent Wannier functions (A). The other two
  bands, lying below $\epsilon_F$ are spanned by another set of two
  equivalent Wannier functions (B), respecting the crystal symmetry.

We would like to emphasize that these orbitals are far from atomic
like. We estimate their spread from the square root of the spread
functional of Wannier90, which yields $5.2$\,\AA{} for orbital type (A), and
$4.4$\,\AA{} for orbital type (B). 
We also want to note that the Wannier functions are not centered on a
Mo site. Instead, 
\fig{fig:latticeWannier} clearly shows that 
the centers are located in the middle of a bond between two Mo sites.
For type (A), one orbital has its center between atoms Mo$_1$ and
Mo$_4^\prime$, the other between Mo$_1^\prime$ and Mo$_4$, resp. In that
sense, these Wannier orbitals can be regarded as bond-centered
molecular-like orbitals.

The origin of the large spread in real space is the very limited
number of bands that are taken into account in the Wannier
construction scheme. Taking all $d$ orbitals of the
Mo$_1$, Mo$_1^\prime$, Mo$_4$, and Mo$_4^\prime$ atoms as well as the
bridging oxygen $p$-orbitals into account would of course 
result in much better localization. However, the Hamiltonian
then describes many bands, and not only the most important four bands in the vicinity of
$\epsilon_F$. A similar effect can be observed
for instance in the
construction of the one-band model in cuprate superconductors. Also there,
taking only the $d_{x^2-y^2}$ orbital into the construction results in
quite large Wannier orbitals with long tails.~\cite{andersen}

Concerning the electron charge in the Wannier orbitals we find that
orbitals of type (A) are {\em half-filled}, whereas orbitals of
type (B) are identified as (almost) filled. For lithium-vacant purple
bronze, we find a total occupation of $\approx5.8$ electrons in
these four bands, since there are two lithium ions in the unit cell,
each contributing $\approx0.1$ hole doping. In the remainder of the
paper, we will therefore use for all discussions an average filling of
the four bands of $\langle n \rangle = 1.44$.~\cite{footnote7} 

% DOWNFOLDING RESULTS
Specifically, the down-folding procedure yields the matrix elements
of a single-particle Hamiltonian~\cite{footnote10}
\begin{align}
 \nonumber H_{\text{Wannier},\alpha\beta}(\mathbf{k}) &=
 \bra{\omega_\alpha}\hat{\mathcal{H}}_{\text{Wannier}}(\mathbf{k})\ket{\omega_\beta}\\ 
&=\sum\limits_\mathbf{\delta R} e^{-i\mathbf{k}\cdot\mathbf{\delta R}} M_{\mathbf{R}\alpha\mathbf{R'}\beta}\,\mbox{,}
\label{eq:Hwannier}
\end{align}
in the four-orbital Wannier space $\alpha,\beta=\{A,A',B,B'\}$ where
the sum runs over all lattice translations $\mathbf{\delta R} =
(\mathbf{R}-\mathbf{R'})$ and the crystal momentum $\mathbf{k}$ is
defined in the first BZ.~\cite{footnote2} 

% MODEL DETAILS
Our model \eq{eq:Hwannier} consists of two filled electronic
orbitals, type (B), slightly below the Fermi energy $\epsilon_F$
($M_{\mathbf{0}B\mathbf{0}B}=M_{\mathbf{0}B'\mathbf{0}B'}=-0.423$\,eV),
as well as two
half-filled ones, type (A) crossing the Fermi energy $\epsilon_F$
($M_{\mathbf{0}A\mathbf{0}A}=M_{\mathbf{0}A'\mathbf{0}A'}=0.005$\,eV). The
largest energy scale for the hopping matrix elements is the
nearest-neighbor hopping
along the 
$b$ direction of orbitals of type (A) which is
$t_{\text{max}}\approx-0.35\,$eV. 

This $4\times 4$ noninteracting Wannier Hamiltonian can easily be
diagonalized by numerical means. Its band structure and DOS are plotted on
top of the LDA results in \fig{fig:bsDos}. The Wannier
DOS has been calculated from $N_{\text{1.BZ}}=48^3$ $k$-points in the
first BZ, using a numerical broadening of $0^+=0.086\,t_{\text{max}}$. We
obtain very good agreement except for the upper band edges, where
the accuracy is influenced by the entanglement of the bands in this
energy region. We find a total bandwidth of the four bands in
the vicinity of $\epsilon_F$ of $W\approx1.57\,$eV and two Fermi
velocities of $v_{F,1}\approx 1.16\cdot10^5\,\frac{\text{m}}{\text{s}}$ and $v_{F,2}\approx 1.06\cdot10^5\,\frac{\text{m}}{\text{s}}$. Note that the Fermi velocity is pointing along the $b$ direction, while the other components are three orders of magnitude smaller. The Fermi surface (see 
\fig{fig:FS} (center)) is also reproduced very accurately by the
Wannier model.

% MODEL STRUCTURE
\begin{figure}[ht]
\includegraphics[width=0.45\textwidth]{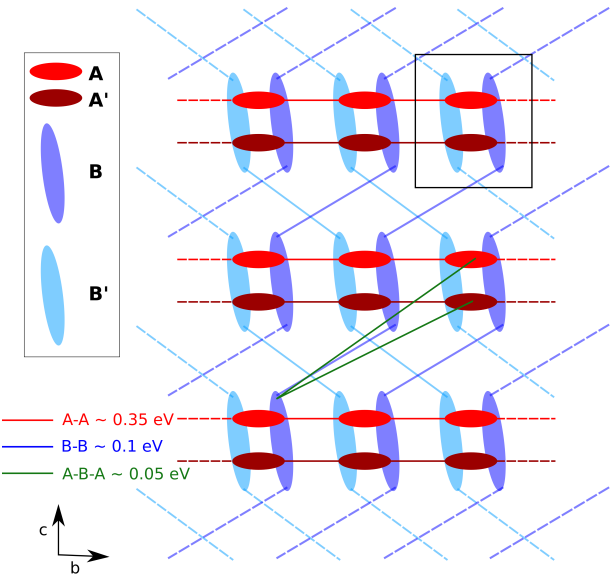} 
\caption{(Color online) Visualization of the full effective Wannier
  Hamiltonian \eq{eq:Hwannier}. A schematic drawing
  of the most dominant hopping processes is presented: type A orbitals
  (red), type B orbitals (blue). Lines denote the dominant hopping
  paths, and the black square marks the size of the unit cell.
}   
\label{fig:hamiltonianVis}
\end{figure}

%\subsection{Effective model analysis by projection}
\subsection{Effective inter-chain coupling}\label{ssec:minimalmodel}
% WE PREPARE AN EVEN SIMPLER MODEL
The Wannier model \eq{eq:Hwannier} consists of numerous single-particle hopping terms between four Wannier orbitals in a three dimensional crystal. 
Many of the terms of the Wannier model are orders of magnitude 
smaller than the dominant hopping process of type (A) orbitals along
$b$ direction with $t_{AA}\approx-0.35\,$eV. For instance, all
intra-unit-cell hybridizations are negligibly small (of order
$10^{-4}\,$eV). This includes direct hopping $t_{AA'}$ between adjacent
chains. The reason for this is that the two orbitals type (A) and (A')
are aligned parallel to each other in the unit cell, with negligible
overlap. The hybridization perpendicular to the chains, which is
responsible for the dispersion in perpendicular direction, is predominantly
mediated through the type (B) orbitals in an (A-B-A) or (A-B-A')
fashion, see \fig{fig:hamiltonianVis}. In this section, 
we 
derive a two-dimensional model in the $b$-$c$ plane consisting of two degenerate half-filled chains that  
comprises the fundamental model. The indirect hopping results only in a small effective hopping between the chains, which we estimate perturbatively.

% CRUDE WAY: BY NEGLECT OF MATRIX ELEMENTS ``ZEROTH ORDER''
The starting point for perturbation theory is a Hamiltonian, where
orbitals of type $A$ and type $B$ are decoupled. For this purpose we
define a complete set of projection operators projected Hamiltonians $\hat{\mathcal{H}}_{\alpha \alpha} =
\hat{P}_\alpha\hat{\mathcal{H}}\hat{P}_\alpha$ on the type
$\alpha=\{(A),(B)\}$ orbitals. In zero-th order approximation, the
hybridization terms are set to zero,
$\hat{\mathcal{H}}_{AB}=\hat{\mathcal{H}}_{BA}=0$. For our Wannier model this corresponds to neglecting
those matrix elements which are less than ten percent of the largest occurring hopping
energy $|t_{\text{max}}|=0.35\,$eV and leads to a Hamiltonian 
{ \footnotesize
\begin{align}
 H_{\text{reduced}}&= 
\begin{pmatrix} 
    t_{AA} 2\cos{(k_b b)} & 0 \\
    0 & \epsilon_{B} +t_{BB} 2 \cos{(k_b b \pm k_cc)} \\ 
\end{pmatrix}\mbox{,}
\label{eq:minimal}
\end{align}
}                                 
with $t_{AA}=-0.35$\,eV, $t_{BB}=-0.11\,$eV and
$\epsilon_{B}=-0.45\,$eV accompanied by the rigid band shift of
$\mu=-0.03\,$eV. One has to keep in mind that both bands A and B are doubly degenerate. We will refer to this Hamiltonian as {\em reduced model} throughout
this work.

Note that the two type (B) orbitals disperse in orthogonal
diagonals. The bands crossing the Fermi energy arise due to
the two degenerate type (A) orbitals which now represent isolated, one-dimensional chains
dispersing in $b$ direction. Due to the missing hybridization between
A and B orbitals, fine features of the perpendicular ($c$-direction)
dispersion are not reproduced. Nevertheless, despite its simplicity,
the band structure and density of states (see \fig{fig:bsDos}) are still described very well.
Data in the figure have been obtained using $N_{\text{1.BZ}}=48^3$
$k$-points in the first BZ and a numerical broadening of
$0^+=0.086\,t_{\text{max}}$ for the evaluation of the DOS.
We find a bandwidth of the two bands in the vicinity 
of $\epsilon_F$ of $W\approx1.4$\,eV and a Fermi velocity of
$v_{F}\approx 0.93\cdot10^5\,\frac{\text{m}}{\text{s}}$.

In order to estimate the effective inter-chain coupling we treat the indirect 
(A-B-A$^{(}$'$^{)}$) hoppings in second-order perturbation theory. For that purpose we project the full four-orbital Wannier model \eq{eq:Hwannier} onto the type (A) bands,~\cite{footnote8}
\begin{align*}
 \hat{\widetilde{\mathcal{H}}}_{AA} &= \hat{\mathcal{H}}_{AA} +\hat{\mathcal{H}}_{AB}\left(\omega-\hat{\mathcal{H}}_{BB}\right)^{-1}\hat{\mathcal{H}}_{BA}\,\mbox{.}
\end{align*}
Upon approximating $\omega$ by the bare eigen-energies of
$\hat{\mathcal{H}}_{AA}$, we arrive at a two-orbital model which
reproduces the band dispersions of the two bands crossing the Fermi
energy (not shown). We note in passing that this two-orbital model
can also be obtained by a Wannier construction where the basis is
restricting to bands of type A alone.

Keeping the number of hopping terms low, we now perform a fit of 
$\hat{\widetilde{\mathcal{H}}}_{AA}(\mathbf{k})$ with
a Hamiltonian that contains perpendicular hopping in addition to the
terms of the (A) orbitals of the reduced model (\eq{eq:minimal}) respecting the symmetry of the lattice. 
In particular, we choose for the perpendicular hopping both
intra-chain ($A$-$A$) terms as well as inter-chain ($A$-$A'$) terms. 
The $\chi^2$ fit is done using $20^2$ $k$-points on an equidistant
grid in one fourth of the reciprocal $b$-$c$ plane ($k_a\approx0$)
plus $3\cdot 32$ $k$-points on the standard path Y-G-X-M. The only relevant perpendicular hopping processes given by this
procedure are nearest-neighbor inter-chain terms of the order of
$t_{AA'}\approx-0.005\,$eV, as well as nearest neighbor intra-chain
terms of the order of $t_{AA}\approx-0.02\,$eV. The hopping in $b$ direction only slightly renormalizes to $t_{AA/A'A'}\approx-0.37\,$eV accompanied by an on-site shift of $\epsilon_{A/A'}=-0.01\,$eV.

Thus, we find an intuitive two orbital model that consists of two
chains dispersing  in $b$ direction with nearest neighbor
perpendicular hoppings of type $A$-$A$ and $A'$-$A'$ which are one order 
of magnitude smaller than the hopping in $b$ direction. The direct effective
hopping of type $A$-$A'$ between the two chains within one unit cell is
again one order of magnitude smaller. This small effective coupling
explains the robust one-dimensionality of the compound. Our calculated values are in good agreement with those discussed in
\tcite{PhysRevB.86.075147}. 

We want to stress here that only in this section fitting of parameters was
performed, in order to estimate the effective perpendicular hopping
using only a few parameters. In all other parts of this work, only {\em ab-initio} calculated hopping integrals are used.

\section{Anisotropic conductivity}\label{sec:anisotropicconductivity}
We augment our discussion of the electronic structure by computing the linear response transport and comparing it to experiments. 

% EXPECTED STRUCTURE
The conductivity tensor of \LIMOO{} consists of three independent
diagonal $\sigma_{a},\sigma_{b},\sigma_{c}$ entries as well as one
non-zero off-diagonal element $\sigma_{bc}=\sigma_{cb}$ (see \app~\ref{app:appLRT}).  
\begin{table}
\begin{tabular}{| c | c | c | c | c |}
\hline
Ref. & $\rho_a$    & $\rho_b$    & $\rho_c$    & ratio \\
\hline  
     & m$\Omega$cm & m$\Omega$cm & m$\Omega$cm &       \\
\hline   
\hline                       
\tcite{Greenblatt1984671} & $2470$ & $9.5$ & - & 260:1:-\\
\tcite{0295-5075-89-6-67010} & $64.5$ & $16$ & $854$ & 4.5:1:50\\
\tcite{PhysRevB.69.085120} & - & $1.7$ & - & -:1:-\\
\tcite{PhysRevB.76.233105} & $110(40)$ & $19(1)$ & $47(5)$ & 6(2):1:2.5(4)\\
\tcite{PhysRevLett.108.187003} & $30$ & $0.4$ & $600$ & 80:1:1600\\
\tcite{PhysRevLett.102.206602} & - & $0.4$ & - & 100:1:$>$100\\
\tcite{NatComm.2.396} & - & - & - & 100:1:-\\
full Wannier model & $\approx430\gamma$ & $\approx1.8\gamma$ & $\approx600\gamma$ & 240:1:330\\
reduced model & - & $\approx2\gamma$ & - & -:1:-\\
\hline  
\end{tabular}
\caption{Collected data for the anisotropic resistivity at $T=300\,$K
  and our low temperature theoretical results for small scattering $\gamma[$eV$]\sim0.05\,$eV. Data from \tcite{Greenblatt1984671}, \tcite{0295-5075-89-6-67010} and \tcite{PhysRevB.69.085120} was obtained via four-point measurements, \tcite{PhysRevB.76.233105} and \tcite{PhysRevLett.108.187003} report results using the Montgomery method, \tcite{PhysRevLett.102.206602} measurements are based on magneto resistance and in \tcite{NatComm.2.396} a Hall experiment was carried out.}
\label{tab:conductivity}
\end{table}
%
% Literature
Literature provides values for the anisotropic resistivity at room
temperature ($300\,$K) and zero magnetic field using several
experimental techniques. We summarized the reported data in \tab{tab:conductivity} which all
agree on a highly anisotropic resistivity. The ratio between the
diagonal elements of the resistivity tensor, however, strongly
disagrees in between the individual measurements. In particular
$\rho_a:\rho_b$ differs by a factor of $\approx 60$ while
$\rho_b:\rho_c$ differs even by a factor of $\approx 640$ from the lowest
to the highest anisotropy found in experiments. These discrepancies
are often attributed to experimental challenges when measuring the
resistivity of strongly anisotropic small samples. To our knowledge
we present the first theoretical study of the conductivity of \LIMOO{}. 

\subsection{Conductivity of the reduced model}\label{sec:MMConductivity}
The reduced model introduced in \eq{eq:minimal} consists of $N_{\text{band}}=2$ degenerate bands (type A) crossing the Fermi energy dispersing only in $b$ direction with velocity $v_{b}^{AA}(\mathbf{k})= -\frac{2t_{AA}}{\hbar}b\sin{(k_b b)}$ (\eq{eq:vel}). In this case of diagonal velocities and spectral functions, the conductivity (see \app~\ref{ssec:lrt}) becomes 
\begin{align*}
\sigma_{bb} &= \frac{16e^2t_{AA}^2}{\hbar}\frac{b^2}{ac}\int_{-\infty}^{\infty}d\omega \frac{\beta}{2(1+\cosh{(\beta(\omega-\mu))})}\\
&\times \int_0^{\frac{\pi}{b}}dk_b \sin^2{(k_b b)}(\frac{\gamma}{\pi})^2\frac{1}{((\omega-2t_{AA}\cos{(k_b b)})^2-\gamma^2)^2}\,\mbox{,}
\end{align*}
where we introduced a phenomenological scattering $\gamma\sim|\Im\text{m}(\Sigma(\omega=\mu))|$ in the Lorentzian-shaped spectral function.
In the low-temperature small-scattering limit we find
\begin{align*}
\sigma_{bb} &= \frac{16e^2t_{AA}^2}{\hbar}\frac{b^2}{ac}\int_{-\infty}^{\infty}d\omega \delta(\omega-\mu)\\
&\times \int_0^{\frac{\pi}{b}}dk_b\sin^2{(k_b b)}\frac{\delta(\omega-2t_{AA}\cos{(k_b b)})}{2\pi\gamma}\,\mbox{,}
\end{align*}
which evaluates to
\begin{align*}
\sigma_{bb} &= \frac{4e^2}{h\gamma}\frac{b}{ac}\sqrt{(2t_{AA})^2-\mu^2}
&\stackrel{\mu= 0}{\approx} N_{\text{spin}}N_{\text{band}}\frac{D}{R_K\gamma}\frac{b}{ac}\,\mbox{,}
\end{align*}
with $D=2t_{AA}$ and $R_K=\frac{h}{e^2}$ the von-Klitzing constant. Using this
expression we find for the resistivity
$\rho_{bb}=\frac{1}{\sigma_{bb}}\approx
2\gamma[$eV$]\,$m$\Omega$cm. Considering a reasonable mean-free path
$d$ of the order of a unit-cell length and using the calculated Fermi
velocity of $\approx 10^5\,\frac{\text{m}}{\text{s}}$ we can estimate a
scattering of $\gamma[$eV$]=\frac{0.658}{d[\AA]}\approx 0.05\,$eV which implies a
resistivity of $\rho_{bb}\approx
0.1\,$m$\Omega$cm. 

% RESULTS
\subsection{Conductivity anisotropy}\label{ssec:cwm}
The reduced model is limited to transport in $b$ direction. To study the high transport anisotropy suggested by experiments, we calculate the conductivity tensor of the full four-orbital Wannier model. We evaluate \eq{eq:A0} numerically at $T=4.2\,$K and for small scattering $\gamma$ (for details see \app~\ref{app:anisotropicConductivity}). For the resistivity we obtain $\rho_a\approx430\gamma[$eV$]$\,m$\Omega$cm, $\rho_b\approx1.8\gamma[$eV$]$\,m$\Omega$cm, and
$\rho_c\approx600\gamma[$eV$]$\,m$\Omega$cm
($\rho_a:\rho_b:\rho_c\sim240:1:330$). Note that the $b$ axis
resistivity in the strictly one-dimensional model \eq{eq:minimal}
is only $\approx 10\%$ larger than the resistivity in the four-orbital
model, which means that the 
reduced model yields already a quite accurate description of the $b$ axis transport. The $a$ and $c$ axis resistivities are
roughly two orders of magnitude larger than in $b$ direction,
compatible with experimental data. Using the same phenomenological
scattering $\gamma\approx0.05$ as motivated in the previous section we
obtain $\rho_a\approx20\,$m$\Omega$cm, $\rho_b\approx0.1\,$m$\Omega$cm
and $\rho_c\approx30\,$m$\Omega$cm. The resistivity ratio of
$\rho_a/\rho_b\sim240$ does compare best to the experimental value
$260$ as obtained in \tcite{Greenblatt1984671} (see \tab{tab:conductivity}). 

\section{Correlated electronic structure}\label{sec:correlatedelectronicstructure}
% GENERAL INTERACTING MODEL
The obtained Wannier model is close to being half-filled which
indicates that
the local part of the Coulomb
interactions is most important. The effects of off-diagonal
Coulomb interactions are small and further discussed in 
\app~\ref{app:extVCA}. In the following, we focus on local
electron-electron interactions of density-density type, which are
added to the {\em ab-initio} tight-binding Wannier model 
\begin{align}
 \hat{\mathcal{H}} &= \hat{\mathcal{H}}_{\text{Wannier}} + \hat{\mathcal{H}}_{\text{int}}\,\mbox{,}
\label{eq:H}
\end{align}
where the single-particle part $\hat{\mathcal{H}}_{\text{Wannier}}$ is defined in \eq{eq:Hwannier} and
\begin{align}
 \hat{\mathcal{H}}_{\text{int}} &= \sum\limits_{\mathbf{R}} \sum\limits_{\alpha}\,U_\alpha\hat{n}_{\mathbf{R}\alpha\uparrow}\hat{n}_{\mathbf{R}\alpha\downarrow}\label{eq:Hint1}\,\mbox{,}
\end{align}
where $\hat{n}_{\mathbf{R}\alpha\sigma}$ is the particle number operator for Wannier orbital $\alpha=\{A,A',B,B'\}$ and spin $\sigma=\{\uparrow,\downarrow\}$ in unit cell $\mathbf{R}$. In order to treat the different band fillings in the model properly we employ a simple double counting correction\cite{Ryndyk2012} in $\hat{\mathcal{H}}_{\text{Wannier}}$,
\begin{align}
M_{\mathbf{0}\alpha\mathbf{0}\alpha}^{+\text{DC}}&= M_{\mathbf{0}\alpha\mathbf{0}\alpha}- U_{\alpha}\langle n_{\mathbf{0}\alpha} \rangle_{\text{Wannier}}\,\mbox{,} 
\label{eq:dc}
\end{align}
where the densities $\langle n \rangle_{\text{Wannier}}$ are taken from the noninteracting Wannier model. Adding interactions we furthermore set
the chemical potential $\mu$ such that the average filling of
electrons in the system is at its physical value $\langle n \rangle
\approx 1.44$.  

% PLAN OF ATTACK
This interacting theory is challenging to solve, even more so because the expected results hint to low dimensional physics which can promote non-local self-energy effects. We employ two complementary techniques to study on a first, more qualitative level, the effect of interactions. First, we apply the VCA, which contains non-local contributions to the self-energy $\Sigma$. Second, we augment these results by a DMFT calculation which neglects contributions of non-local self-energy terms, but is superior in the treatment of the low-energy quasi-particle resonance.
\begin{figure*}
\includegraphics[width=0.99\textwidth]{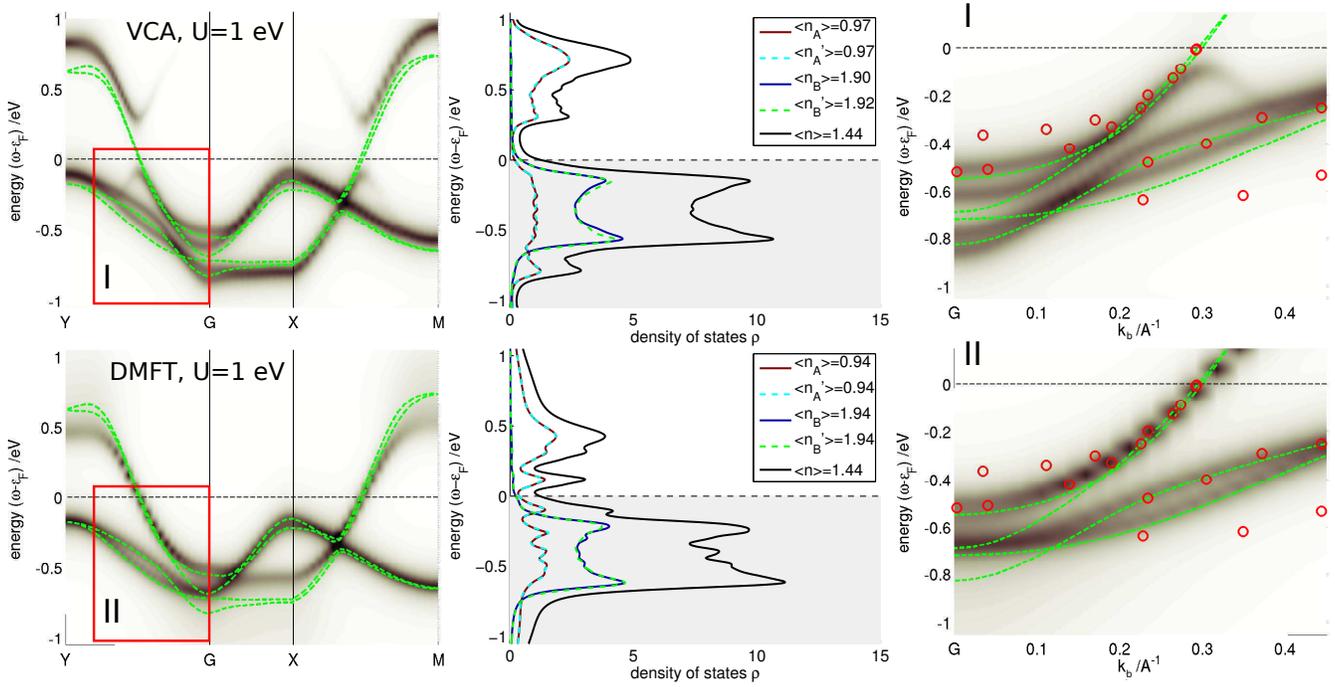}
\caption{(Color online) Spectral function and DOS of the
  interacting model. Top row: VCA data for $U=1\,$eV using
  eight-orbital clusters. Bottom row: DMFT data for $U=1\,$eV. Left:
  Spectral function plotted along a two dimensional path in the
  reciprocal $b$-$c$ plane. The noninteracting dispersion is plotted
  on top (dashed-green). Center: Orbitally resolved density of
  states. Right: Zoom to the spectral function in the respective red
  rectangle compared to ARPES data from \tcite{PhysRevLett.103.136401}
  which are indicated as red circles.}  
\label{fig:interacting}
\end{figure*}

\subsection{Variational Cluster Approach}\label{ssec:eVCA}
The VCA~\cite{potthoff_variational_2003} is a quantum many-body cluster method which is capable of treating short range correlations exactly.~\cite{potthoff_variational_2003,PhysRevB.70.245110,PhysRevB.70.235107,senechal_2009} The given lattice Hamiltonian is partitioned into a cluster- and an inter cluster-Hamiltonian $\hat{\mathcal{H}}=\hat{\mathcal{H}}^{\text{cl}} + \hat{\mathcal{H}}^{\text{inter}}$, where only single-particle terms are allowed in the inter cluster part (for details see \app~\ref{app:extVCA}). Clusters consist of one or more unit cells and are chosen so that their single-particle Green's function $g_{ml}^\sigma(z)$ can be obtained exactly. We use clusters consisting of two unit cells in $b$ direction ($L_\mathcal{C}=8$) to capture at least the most basic non-local self-energy effects on the quasi one-dimensional chains which enable signatures of a possible spin-charge separation.~\cite{PhysRevB.72.155110} In this work we employ a numerical Band Lanczos scheme and the Q-matrix formalism to 
obtain $g_{ml}^\sigma(z)$.~\cite{PhysRevB.74.235117} The CPT approximation to the
single-particle Green's function of the full system $G^{-1}(z)$ is given within first order strong coupling perturbation theory by~\cite{gros_cluster_1993, senechal_spectral_2000} 
\begin{align}
 G^{-1}(z) = g^{-1}(z)-H^{\text{inter}}\,\mbox{,}
\label{eq:CPT}
\end{align}
where $H^{\text{inter}}$ are the matrix elements of the inter cluster Hamiltonian in the basis of cluster orbitals. If the cluster is larger than the actual unit cell of the crystal, we use a Green's function periodization prescription to project on the original unit cell $G_{\alpha\beta}(z)$~\cite{senechal_2009}. 

%VCA
Within VCA, \eq{eq:CPT} is evaluated at the stationary point of the generalized grand potential $\Omega[\boldsymbol{\Sigma}]$ (for fermions at zero temperature) which is available from $G$ and $g$.~\cite{potthoff_variational_2003} The grand potential is parametrized by the VCA variational parameters $\boldsymbol{\Delta}$ which are fixed by the VCA condition~\cite{potthoff_variational_2003} 
\begin{equation}
\boldsymbol{\nabla}_{\boldsymbol{\Delta}}\Omega(\boldsymbol{\Delta})
 \stackrel{!}{=}\mathbf{0}\,\mbox{.} 
\label{eq:omegavca}
\end{equation}
The VCA improves the CPT ($\mathbf{\Delta}\equiv \mathbf{0}$) approximation (which is to approximate the self-energy $\Sigma_G$ of the full system by the self-energy of the cluster $\Sigma_g$) by adding flexibility to the cluster self-energy in terms of variational parameters $\mathbf{\Delta}$. We consider the on-site energies of the four Wannier orbitals as independent variational parameters $\boldsymbol{\Delta} = \{\Delta_{\epsilon_A},\Delta_{\epsilon_{A'}},\Delta_{\epsilon_B},\Delta_{\epsilon_{B'}}\}$ (which implicitly includes an overall shift of the chemical potential of the cluster) and use $N_{\text{1.BZ}}=32^3$ $k$-points in the irreducible BZ for the evaluation of \eq{eq:omegavca}.

% CHOICE OF CLUSTERS AND VCA PARAMETERS AND LIMITATIONS
Advantages of the VCA are that i) it is exact in the noninteracting system, ii) the approximation is systematically improvable by enlarging cluster sizes, iii) or increasing the number of variational parameters $\mathbf{\Delta}$ and iv) it is possible to work directly in the real energy domain as well as in Matsubara space. VCA on small clusters is inherently biased towards the insulating state, therefore we expect to overestimate a possible Mott gap (see \app~\ref{app:VCAclusterSize} for a discussion). 
\begin{figure}
\includegraphics[width=0.49\textwidth]{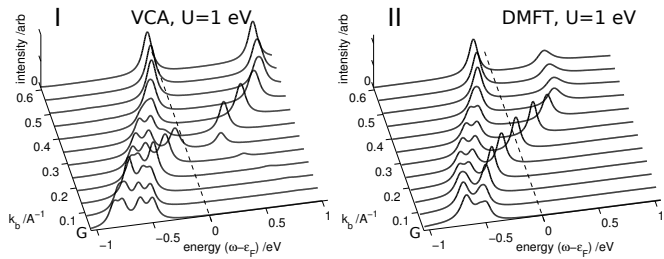}
\caption{(Color online) Cuts through the spectral function along the
  $b$ axis. The parameters and labels (I, II) correspond to those in
  \fig{fig:interacting}.} 
\label{fig:interactingcuts}
\end{figure}

\subsection{Dynamical Mean Field Theory}\label{ssec:DMFT}
% DMFT
A complementary approach that neglects non-local effects but describes the local dynamical quantum fluctuations better is the DMFT.~\cite{RevModPhys.68.13} Within this theory the interacting lattice problem is mapped on a self-consistent four-orbital impurity model coupled to an infinite electronic bath. The DMFT approximation is to assume a
momentum-independent self-energy of the original model
$\Sigma_{\alpha\beta}$ 
\begin{align*}
 \Sigma_{\alpha\beta}(i\omega,\mathbf{k}) &\stackrel{!}{=} \mathcal{S}_{\alpha\beta}(i\omega)\,\mbox{,}
\end{align*}
where $\mathcal{S}_{\alpha\beta}$ is the local self-energy generated by the auxiliary quantum impurity system.

% IMPURITY SOLVER
As impurity solver we use the continuous time Quantum Monte Carlo (CT-QMC) code of the TRIQS~\cite{triqs_project} toolkit and its implementation of the hybridization expansion
(CT-HYB)~\cite{triqs_ctqmc_solver_werner1, PhysRevB.74.155107} algorithm using Legendre polynomials.~\cite{PhysRevB.84.075145} This sign-problem free method works in Matsubara space and provides statistically exact and reliable results even at very low temperatures.~\cite{RevModPhys.83.349} We used a low temperature of $\beta=150\,$eV$^{-1 }$ and a $k$-mesh of $N_{\text{1.BZ}}=800$. The imaginary time data is continued to the real frequency axis using a parallel tempering analytic continuation method.~\cite{Beach2004}

% LIMITATIONS
Different to VCA, the DMFT as applied here neglects non-local correlations. On the other hand, it treats the local dynamical quantum fluctuations accurately. 

\begin{figure*}[t]
\includegraphics[width=0.99\textwidth]{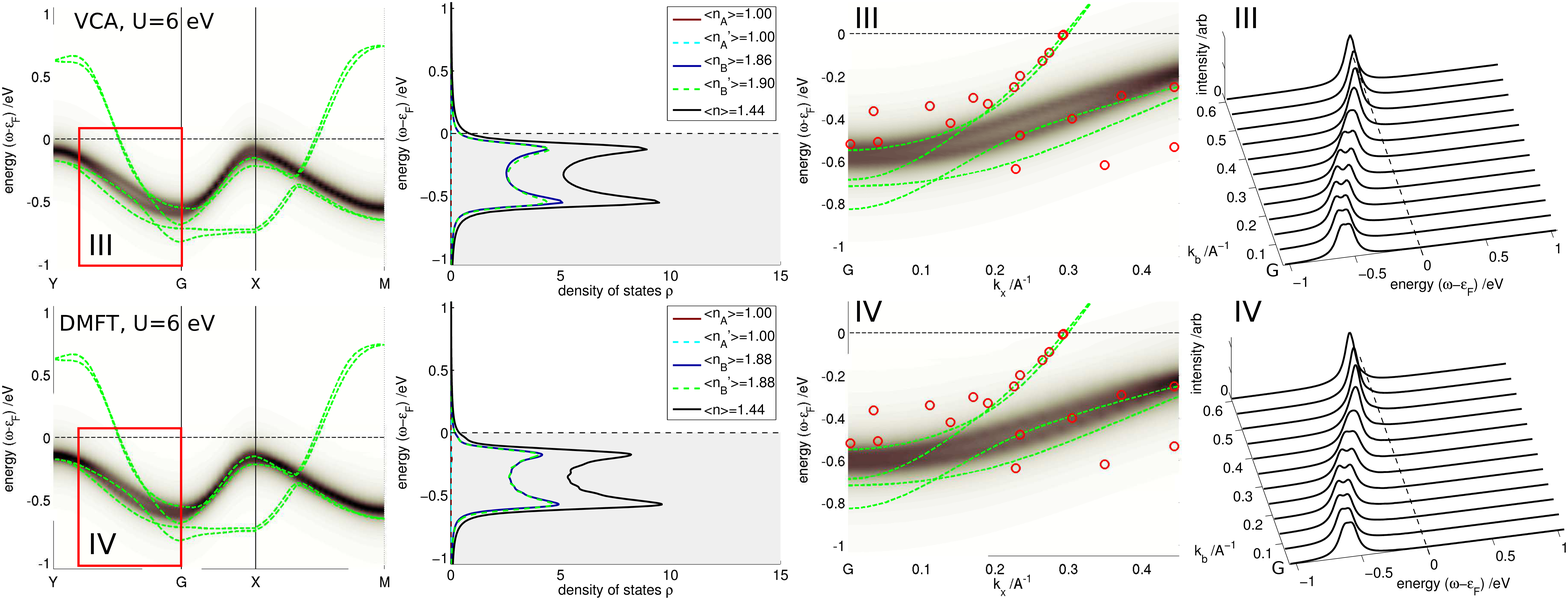}
\caption{(Color online) Spectral function and DOS of the
  interacting model for high values of on-site interaction strength as
  one would expect for atomic like molybdenum $d$ orbitals:
  $U=6\,$eV. For legends, arrangement of the subplots and color coding
  see \fig{fig:interacting}. On the far right we show cuts through the
  spectral function as in \fig{fig:interactingcuts}. } 
\label{fig:highU}
\end{figure*}

\subsection{Discussion of the interacting dynamics}\label{ssec:Interacting Dynamics}
For models based on {\em atomic} orbitals, constrained
LDA calculations suggest an on-site interaction for the atomic $d$ Mo
orbitals of $U\approx6.4$\,eV and a nearest-neighbor interaction of
$V\approx0.2\,$eV~\cite{PhysRevB.74.045117} in \LIMOO, while the bulk
Mo value for the on-site interaction is $U\approx3.8\,$eV.~\cite{PhysRevB.86.075147} As we will discuss below,
in our model these larger interaction values $U$, which have been
proposed and used for model calculations,~\cite{PhysRevB.85.235128,PhysRevB.86.075147} 
do not give results in accordance with experimental data. Obtaining the interaction parameters in an {\em ab-inito} way by,
e.g. the constrained Random Phase Approximation (cRPA),~\cite{PhysRevB.80.155134,JPSJ.79.112001} would be highly desirable, but is beyond our present computational capabilities due to the very large unit cell of the system.

In the following, based on physical arguments, we will nevertheless argue that a
moderate value of $U$ is appropriate for our model. We use uniform {\em on-site} interactions $U_\alpha=U$ only and
estimate the magnitude of the interaction strength to be of the order
of a few $t_{\text{max}}$.
The reduced value, compared to the atomic one, can be motivated by i) the large spread of the orbitals,~\cite{Ferber2012}  and ii) the effective screening of other Mo 4d states near the Fermi level. We want to remind the
reader that we are not dealing with atomic orbitals (where for
molybdenum the  interaction $U$ could be of the order of several
electron volts) but with extended, even molecular-like, orbitals (see
\fig{fig:latticeWannier}). 

Let us start the discussion using interaction values of the order of
the band width, i.e. using $U=1$\,eV. We show results for the
single-particle spectrum and orbitally resolved DOS 
of the interacting model in
\fig{fig:interacting} (left and center) and
\fig{fig:interactingcuts}.~\cite{footnote5} We used a Lorentzian broadening of
$0^+=0.05$\,eV for plotting the spectral functions, as well as
$0^+=0.025$\,eV for plotting the DOS. As discussed in the
previous section, the VCA is biased towards an insulating solution (see also \app~\ref{app:VCAclusterSize}),
that is why there is a small gap in the conduction band visible in the
spectral function, which is not seen in the DMFT results.

Comparing the single-particle dynamics to recent experiments (ARPES data from
~\tcite{PhysRevLett.103.136401} and ~\tcite{PhysRevLett.96.196403}) we
find very good agreement for the bands at the Fermi energy
(\fig{fig:interacting} (right)). The renormalization of the effective
mass of the half-filled orbitals, calculated from the DMFT self
energy, is $m\approx 1.2 m_0$, where $m_0$ is the LDA band mass.  

Regarding the bands crossing the Fermi energy, their slope
improves in VCA/DMFT with respect to the LDA data, and compares well
with the measured excitations in ARPES
experiments.\cite{PhysRevLett.103.136401} Note that the up-most branch
provides only a very weak signal in the ARPES data as compared to the
lower branch. In our DMFT calculation, the electronic correlations
suppress the hybridizations between chains (A) and (A'), making them
equivalent. This leads to only one dispersing feature crossing the
Fermi energy, see \fig{fig:interacting} lower right panel. The red
circles at lower binding energy correspond to the shoulder in the
ARPES data, which are very likely due to non-local correlation effects
that are completely neglected in the single-site DMFT approach. A
final statement on the impact of non-locality of the self energy and
spin-charge separation on the single-particle excitations require a
detailed investigation on large systems, which is beyond the scope of
this paper. 

Increasing the interaction value further, for instance to $U=1.5$\,eV
does not change results significantly (left aside the artificial gap
in the VCA calculation). Above a certain limit, however, which is
around $U=2.5$\,eV in our calculations, a Mott gap opens in the
two half-filled bands. An extreme example is using the atomic value
for the interaction, $U=6$\,eV, which is shown in \fig{fig:highU}. The
half-filled bands are in the Mott insulating state, with the spectral
weight transferred to roughly $\pm3\,$eV. The only spectral weight
left close to the Fermi level originates from the two almost filled
orbitals, type (B). This is of course qualitatively different from
experimental results. 

The values of $U$ given here can only be seen as rough estimates to
the actual value, and are by no means {\em ab-initio}. The DMFT overestimates the metallicity of a system, in
particular in low dimensions while the VCA underestimates it. Hence, using different techniques which
are tailored more towards low dimensions, the needed value of $U$ to open a
Mott gap might be even smaller. As has been shown by Chudzinski
\etal,~\cite{PhysRevB.86.075147} the system should be metallic, but
very close to an insulating state. Our observations can be used as guideline in
further studies to determine the value of $U$.
At very large coupling $U=6$\,eV, however, the system as modeled here is
strongly localized, and the insulating state there should be
very robust. This is supported by the fact that both methods, VCA and
DMFT, give indistinguishable results in this (almost) atomic
limit. It clearly shows that taking atomic values for $U$ is inadequate for
the effective model derived in this work.

% MINIMAL MODEL
Let us shortly comment on the effect of
correlations in the reduced model, \se~\ref{ssec:minimalmodel}. There,
the Hamiltonian of the half-filled and the filled bands decouples
exactly, which means that one is left with a standard one-dimensional
(almost) half-filled Hubbard model with nearest-neighbor hopping
only.~\cite{essler_2010} As discussed in \se~\ref{ssec:minimalmodel} and \se~\ref{sec:MMConductivity}, this gives a quite
good description of the dispersion in chain-direction including transport properties. Effects beyond
the one-dimensional Hubbard model can be included using the effective
perpendicular hopping terms as estimated in
\se~\ref{ssec:minimalmodel}. In a recent study on dimensional
crossover\cite{PhysRevLett.109.126404} the critical perpendicular
coupling to enter the regime of one-dimensional physics is $t_p\approx
0.18 t$ at interaction strength $U=3t$. Of course, this value depends
on model details such as frustrated 
hopping and interaction strength. However, since our estimated value for
$t_p$ in \LIMOO{} is significantly smaller than this boundary, we
suggest that this can explain the robustness of 1D physics in this compound. We leave a
more detailed study of the dimensional crossover in \LIMOO{} for
further investigations. 

\section{Conclusions}\label{sec:conclusions}

We have devised a model for the electronic structure of the highly anisotropic low dimensional purple bronze \LIMOO. Starting from {\em ab-initio} calculations, applying Density Functional Theory in the Local Density Approximation, we constructed a four-orbital model based on molybdenum $d$ states in terms of maximally localized Wannier functions. This leads to an effective theory with {\em two filled bands} slightly below and {\em two half-filled bands} crossing the Fermi energy. We obtained an even more elementary effective model with reduced dimensionality consisting of two orbitals only, tailored towards studies of interactions at low energies.

We showed that basic electronic properties of our model are in good
agreement with experimental data and {\em ab-initio}
results. Estimated anisotropic transport coefficient reproduce 
experimental trends. The model enables us to study effects of
many-body correlations. In a first approach we made use of the
(extended) Variational Cluster Approach which takes into account
non-local contributions to the self-energy and Dynamical Mean Field
Theory to study the effects of density-density type electron-electron
interactions. Our results indicate that moderate on-site interactions
(of the order of the band width) are essential while nearest-neighbor
density-density interactions play a minor role. The so obtained
single-particle spectra agree well with recent angle resolved photo
emission experiments. Our study sets some qualitative limits on the
value of the interaction parameters. In particular, we could show that
the values used for atomic-like molybdenum $d$ orbitals are completely
inappropriate for our Wannier model of lithium purple bronze. 

We would like to point out that our model is very different from previously proposed descriptions for \LIMOO{} which were based on atomic orbitals with a comparatively high on-site interaction strength of several electron volts. We suggest that low-energy treatments of this one-dimensional model should start from two half-filled chains with moderate on-site interaction rather than quarter-filled ladder models with high values of on-site interaction strength plus off diagonal interactions.

Our model is intended to serve as a starting point for future studies
of the electronic structure and interactions of \LIMOO{} be it in a
renormalization group - Luttinger liquid or computational many-body
sense. On the latter side it would certainly be interesting to conduct
a more thorough investigation of non-local self-energy effects to
complement our (extended) Variational Cluster Approach
results. In particular, the phenomenon of spin-charge separation
deserves further attention. A theoretical understanding of the phase diagram of the system, i.e.,
the occurrence of superconducting, insulating, or charge ordered states
as function of pressure and temperature, remains a challenging open
question. These studies could be augmented by an {\em ab-inito} calculation of interaction parameters for the Wannier
model by appropriate techniques such as constrained Random Phase Approximation,~\cite{JPSJ.79.112001,PhysRevB.80.155134} making the approach fully {\em ab-initio}. At the moment of
writing, this is not feasible due to the computational complexity.   

\begin{acknowledgments}
We gratefully acknowledge fruitful discussions with Jim W. Allen, Wolfgang von der Linden, Enrico Arrigoni, Christoph Heil, Jernej Mravlje, Fakher Assaad, and in particular Piotr Chudzinski. MN thanks the Forschungszentrum J\"ulich - Autumn School on Correlated Electrons for hospitality. This work was partly supported by the Austrian Science Fund (FWF) P24081-N16 and SFB-ViCoM sub projects F04103, some calculations have been performed on the Vienna Scientific Cluster (VSC). 
\end{acknowledgments}

\appendix

\section{Linear response transport} \label{app:appLRT}
The structure of the conductivity tensor $\sigma_{\alpha\beta}$ of \LIMOO{} follows from the $C_{2h}$ point symmetry as well as physical symmetry
considerations for the conductivity~\cite{Hartmann1984} and can
easily be established by requiring the conductivity tensor to be i)
symmetric for physical reasons and ii) invariant under transformations
with the four lattice point symmetry operations (identity, inversion,
mirror symmetry perpendicular to $a$-axis and two fold rotation around
the $a$-axis) $S_\alpha$: $\sigma=S_\alpha\sigma S_\alpha^T$.
\subsection{Formalism}\label{ssec:lrt}
% ANISOTROPIC CONDUCTIVITY
Following \tcites{PhysRevB.73.035120,PhysRevB.80.085117,Deng2012}, 
linear response transport coefficients can be expressed in terms of kinetic coefficients 
\begin{align}
\nonumber \mathcal{A}^n_{\nu \mu} &=
N_{\text{spin}}\pi\hbar\int\limits_{-\infty}^{\infty}d\omega\,(\beta\omega)^n\\
&\times p_{\text{FD}}(\omega,\mu,\beta)p_{\text{FD}}(-\omega,-\mu,\beta)\Gamma_{\nu
  \mu}(\omega,\omega)\,\mbox{,} 
\label{eq:A0}
\end{align}
where $N_{\text{spin}}=2$ is due to spin degeneracy, the indices
$\nu, \mu=\{a,b,c\}$ denote the real space coordinate system, and we
neglect vertex corrections. The Fermi-Dirac distribution
$p_{\text{FD}}(\omega,\mu,\beta)=\frac{1}{e^{\beta (\omega-\mu)}+1}$ restricts
the interval of integration to $\beta^{-1}\sim k_B T$ around the Fermi
energy $\epsilon_F$ ($k_B$ is Boltzmann's constant, and $T$ and
$\beta$ denote temperature and inverse temperature, resp.). The transport distribution 
\begin{align}
\nonumber  \Gamma_{\nu \mu}(\omega_1,\omega_2) &=
  \frac{1}{V}\frac{1}{N_{\text{1.BZ}}}\sum\limits_{\mathbf{k}\in1.\text{BZ}}\text{Tr}\\
&\times{\left[v_{\nu}(\mathbf{k})
      A(\omega_1,\mathbf{k}) v_{\mu}(\mathbf{k})
      A(\omega_2,\mathbf{k})\right]}\,\mbox{,} 
\label{eq:gammatransport}
\end{align}
($V=abc$ is the unit cell volume) is given in terms of the velocities
\begin{align}
  v_{\nu}^{\alpha \beta}(\mathbf{k}) &= -\frac{\hbar}{m} \bra{\Psi_\alpha(\mathbf{k})}\nabla_\nu\ket{\Psi_\beta(\mathbf{k})}\,\mbox{,}
\label{eq:velocity}
\end{align}
and the spectral function
\begin{align}
  A_{\alpha \beta}(\omega,\mathbf{k}) &= -\frac{1}{\pi}\Im{\text{m}(G^R_{\alpha \beta}(\omega,\mathbf{k}))} \,\mbox{,}
\label{eq:spectral}
\end{align}
which both are matrices in orbital indices $\alpha, \beta =\{A,A',B,B'\}$, which the trace $\text{Tr}$ runs over.

We use velocities $v_{\nu}^{\alpha\beta}(\mathbf{k})$
(\eq{eq:velocity}) in the Peierls approximation (neglecting the gradient of
the Wannier orbital itself leading to a diagonal representation) 
\begin{align}
\nonumber v_{\nu}^{\alpha\beta}(\mathbf{k})&=\frac{1}{\hbar}\Bigg(\bra{\omega_\alpha(\mathbf{k})}\frac{\partial\hat{\mathcal{H}}(\mathbf{k})}{\partial k_\nu}\ket{\omega_\beta(\mathbf{k})}\\
\nonumber &-{\rm \alpha}\left(r_\alpha-r_\beta\right)\bra{\omega_\alpha(\mathbf{k})}\hat{\mathcal{H}}(\mathbf{k})\ket{\omega_\beta(\mathbf{k})}\Bigg)\\
&\approx\frac{1}{\hbar}\frac{\partial E_\alpha(\mathbf{k})}{\partial k_\nu}\delta_{\alpha\beta}
\label{eq:vel}
\end{align}
where the second term in the first expression takes intra-unit cell
processes into account,~\cite{PhysRevB.80.085117} and $r_\alpha$ is the
position of Wannier orbital $\alpha$ inside the unit cell. This term is
neglected in the following because the intra-unit cell hopping
elements are negligibly small. 

The conductivity tensor is 
\begin{align}
 \sigma_{\nu \mu} &= \beta e^2 \mathcal{A}_{\nu \mu}^0\,\mbox{,}
\label{eq:sigma}
\end{align}
with $e$ denoting the electron charge.

\subsection{Details on the evaluation of the anisotropic conductivity}\label{app:anisotropicConductivity}
In this appendix we outline the numerical procedure used for the evaluation of the conductivity tensor \eq{eq:sigma}. These equations contain four additional, auxiliary numerical parameters in which we converge our results: 
i) The spectral function $A_{\alpha\beta}(\omega,\mathbf{k})$ (\eq{eq:spectral}) of the Wannier Hamiltonian is available exactly through the noninteracting retarded single-particle Green's function $G_{\alpha\beta}^R(\omega) = \bra{\omega_\alpha(\mathbf{k})}\frac{1}{\omega+i\gamma-\hat{\mathcal{H}(\mathbf{k})}}\ket{\omega_\beta(\mathbf{k})}$. The broadening $\gamma$ of the spectral function is chosen phenomenologically as described in the man part of the text. For numerical reasons $\gamma$ has to be chosen in accordance with,
ii) the number of $k$-points $N_{\text{1.BZ}}$ in the first BZ for the
sum in \eq{eq:gammatransport}. We obtain converged conductivities for
$N_{\text{1.BZ}}^{1/3}\in[1,67]$ to within a relative error of
$10^{-3}$ using an equidistant grid in the irreducible BZ. We use $\gamma=\{0.1,0.075,0.05,0.025\}$ and rescale all
conductivities with $\gamma$. As a function of $\gamma$, the
resistivities in $a$ and $b$ direction are constant at
$\rho_a\approx(1.8\pm0.05)\gamma$ and $\rho_b\approx(430\pm10)\gamma$
while the resistivity in $c$ direction shows an upwards trend. For our
values of $\gamma$ we find
$\rho_c\approx\{190,300,480,650\}\gamma$. Since the last data point at
$\gamma=0.025$ is already difficult to converge in $N_{\text{1.BZ}}$
we estimate $\rho_c\approx(600\pm150)\gamma$.

\begin{figure*}
\includegraphics[width=0.99\textwidth]{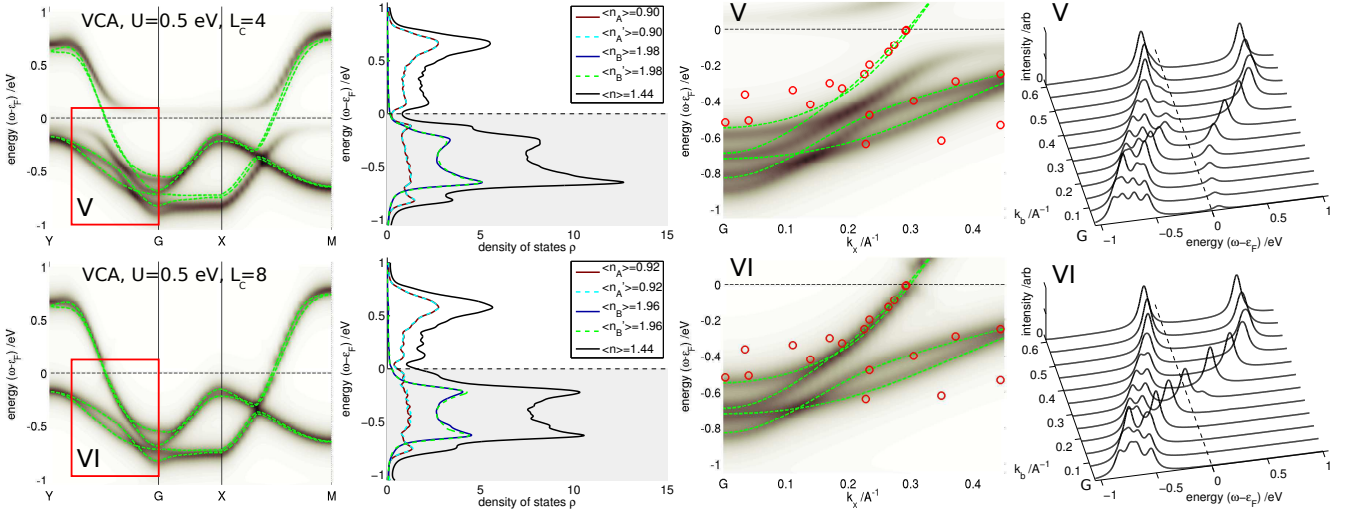}
\caption{(Color online) Comparison of VCA cluster sizes for a moderate
  on-site interaction strength of $U=0.5\,$eV. Top row: VCA data for
  using four-orbital clusters. Bottom row: VCA data for using
  eight-orbital clusters.  For legends, arrangement of the subplots and color coding
  see \fig{fig:highU}.} 
\label{fig:clusterSize}
\end{figure*}

iii) The velocities $v_{\nu}^{\alpha\beta}(\mathbf{k})$
(\eq{eq:velocity}) are obtained by symmetric first order numeric gradient
approximations $ v_{\nu}^{\alpha\beta}(\mathbf{k}) \approx \frac{\delta_{\alpha\beta}}{\hbar}\frac{E_\alpha(\mathbf{k}+\frac{\delta}{2}\mathbf{e}_\nu)-E_\alpha(\mathbf{k}-\frac{\delta}{2}\mathbf{e}_\nu)}{\delta}$ ($\mathbf{e}_\nu$ denotes the unit vector in real space dimension $\nu$).
The parameter of the finite difference scheme for the velocities used is $\delta=10^{-6}$
after finding only negligible changes in a range of
$\delta\in[10^{-8},10^{-3}]$.
iv) For reasons of numerical stability
we evaluate \eq{eq:sigma} at a low, but finite temperature of
$T=4.2\,K$, keeping in mind that $v_{\nu}^{ij}(\mathbf{k})$ and
$A(\omega,\textbf{k})$ have been evaluated for zero temperature. We
find the results to be independent of this choice in a range of
$T\in[1,50]\,K$. In this calculation, at fixed $\gamma,$ the temperature dependence enters
through the Fermi-Dirac distribution only and a small scattering is taken into account through the broadening $\gamma$ in the spectral function.
We checked the numeric procedure on the reduced model where analytic results are known (see main text).

\section{Non-local interactions - extended VCA} \label{app:extVCA}
% eVCA
Here we outline the VCA theory as implemented to obtain the results of the main text including the extensions needed in eVCA to treat non-local Coulomb interactions.\cite{PhysRevB.70.235107} The single-particle part of the full Hamiltonian is readily decomposed into a cluster and an inter cluster part  
\begin{align*}
\hat{\mathcal{H}}_{\text{Wannier}}^{\text{cl}}&=M_{\mathbf{\mathcal{R}}m\mathbf{\mathcal{R}}l}\ket{\omega_m}\bra{\omega_l}\\ 
\hat{\mathcal{H}}_{\text{Wannier}}^{\text{inter}}&=\sum\limits_\mathbf{\delta
  \mathcal{R}} e^{-i\mathbf{k}\cdot\mathbf{\delta \mathcal{R}}}
M_{\mathbf{\mathcal{R}}m\mathbf{\mathcal{R}'}l}\ket{\omega_m}\bra{\omega_l}\,\mbox{,} 
\end{align*}
where indices $m$ and $l$ run over the $L_\mathcal{C}$ orbitals in the cluster $\mathcal{C}$ at superlattice~\cite{senechal_2009} position $\mathbf{\mathcal{R}}$. 

% MANY BODY MEAN FIELD DECOMPOSITION
When off-diagonal interaction terms are non zero, an additional mean-field treatment is needed for those two-particle terms which extend over the cluster boundary.~\cite{PhysRevB.70.235107} This leads to a modified interaction part of the Hamiltonian   
\begin{align*}
 \hat{\mathcal{H}}_{\text{int}} &= \sum\limits_{\mathcal{C}} \left( \hat{\mathcal{H}}_{\text{int}}^{\text{cl}} + 
\hat{\mathcal{H}}_{\text{mf}}^{\text{cl}}(\boldsymbol{\varphi})\right)\\ 
\hat{\mathcal{H}}_{\text{int}}^{\text{cl}}
&=\sum\limits_{m=1}^{L_\mathcal{C}}\,U_m\,\hat{n}_{m\uparrow}\hat{n}_{m\downarrow}+\sum\limits_{\begin{subarray}{c}
    m<l\in \mathcal{C}\\
    \sigma\sigma'\end{subarray}}\,V_{ml}\,\hat{n}_{m\sigma}\hat{n}_{l\sigma'}\\ 
\hat{\mathcal{H}}_{\text{mf}}^{\text{cl}}
(\boldsymbol{\varphi})&=\sum\limits_{m
  l}\,\widetilde{V}_{ml}\left(\sum\limits_{\sigma}\left(\hat{n}_{m\sigma}\varphi_{l}+\hat{n}_{l\sigma}\varphi_{m}\right)-\varphi_{l}\varphi_{m}\right)\,\mbox{,}  
\end{align*}
with on-site interaction strength $U_m$, intra-cluster off-diagonal interactions $V_{ml}$ as well as $\widetilde{V}_{ml}=\sum\limits_{\mathbf{\mathcal{R}}}V_{0m\mathbf{\mathcal{R}}l}$
the interaction elements in the mean-field Hamiltonian. The mean-fields $\boldsymbol{\varphi}$ (taken as spin independent $\varphi_m = \sum\limits_\sigma \langle \hat{n}_{m\sigma}\rangle$ and restricted by lattice symmetry) need to be determined self-consistently. 

% CLUSTER HAMILTONIAN
This allows to write the (interacting) cluster Hamiltonian in the VCA as
\begin{align*}
\hat{\mathcal{H}}^{\text{cl}}(\boldsymbol{\Delta},\boldsymbol{\varphi})
&= \hat{\mathcal{H}}^{\text{cl}}_{\text{Wannier}}(\boldsymbol{\Delta})
+ \hat{\mathcal{H}}^{\text{cl}}_{\text{int}} +
\hat{\mathcal{H}}^{\text{cl}}_{\text{mf}}(\boldsymbol{\varphi})\,\mbox{,} 
\end{align*}
where we introduced the VCA variational parameters~\cite{potthoff_variational_2003,PhysRevB.70.245110} $\boldsymbol{\Delta}$.

% NEAREST-NEIGHBOR THEORY
To study the impact of non-local Coulomb interactions we extend \eq{eq:Hint1} by
\begin{align*}
\hat{\mathcal{H}}_{\text{int,nl}}&= \hat{H}_{\text{int}} + \sum\limits_{\mathbf{R}<\mathbf{R'}}\sum\limits_{\begin{subarray}{c}
    \alpha<\beta\\  \sigma\sigma'\end{subarray}}\,
V_{\mathbf{R}\alpha\mathbf{R'}\beta}\,
\hat{n}_{\mathbf{R}\alpha\sigma}
\hat{n}_{\mathbf{R'}\beta\sigma'}\,\mbox{,}\label{eq:Hint2} 
\end{align*}
which also effects the double counting terms in \eq{eq:dc}
\begin{align*}
M_{\mathbf{0}\alpha\mathbf{0}\alpha}^{+\text{DC,nl}}&= M_{\mathbf{0}\alpha\mathbf{0}\alpha}^{+\text{DC}}-\sum\limits_{\mathbf{R}\gamma}V_{\mathbf{0}\alpha\mathbf{R}\gamma}\langle n_{\mathbf{R}\gamma}\rangle_{\text{Wannier}}\,\mbox{,} 
\end{align*}
where the sum over $(\mathbf{R}, \gamma)$ runs over all bonds connected to orbital $(\mathbf{0}, \alpha)$.
The mean-fields $\boldsymbol{\varphi}$~\cite{PhysRevB.70.235107} which arise due to off-diagonal interaction terms are fixed by the eVCA condition on the generalized grand potential~\cite{footnote4}
\begin{equation*}
\boldsymbol{\nabla}_{\boldsymbol{\Delta},\boldsymbol{\varphi}}\Omega(\boldsymbol{\Delta},\boldsymbol{\varphi})
 \stackrel{!}{=}\mathbf{0}\,\mbox{.} 
\end{equation*}

% NEAREST-NEIGHBOR RESULTS
In order to check the influence of {\em nearest-neighbor density-density interactions} $V_{\mathbf{R}\alpha\mathbf{R}'\beta}$ we did several eVCA calculations with different values within reasonable limits, i.e. below a value of $\approx\frac{U}{2}$. Our calculations show, however, that these interactions $V_{\mathbf{R}\alpha\mathbf{R}'\beta}$ lead only to minor differences compared to results without them. We did not find the system to be susceptible to any charge ordering. For that reason, and also because the precise value of the parameters $V_{\mathbf{R}\alpha\mathbf{R}'\beta}$ is complicated to estimate, all results presented here are calculate with on-site interaction $U_\alpha=U$ only.~\cite{footnote9} Given the band filling factors and the good agreement with ARPES experiments we argue that on-site interactions are sufficient to describe the spectral properties of this system within our approximation. 

\section{VCA cluster size extrapolation} \label{app:VCAclusterSize}
Here we discuss the approximation introduced by choosing eight-orbital
clusters for the VCA procedure. Eight-orbital clusters enable
non-local self-energy effects along the chain direction in the most
basic fashion. The VCA on small cluster sizes is inherently biased
towards the insulating state.~\cite{senechal_2009} In
\fig{fig:clusterSize} we show the behavior of the results when going
from one-unit cell clusters $L_{\mathcal{C}}=4$ to two-unit cell
clusters in $b$ direction $L_{\mathcal{C}}=8$. For the same
interaction strength the $L_{\mathcal{C}}=4$ calculation clearly shows
a pronounced Mott gap in the (A) type orbitals while it is still
absent in the $L_{\mathcal{C}}=8$ calculation. All other basic
features are comparable. For numerical reasons we can not go to larger
cluster sizes. Nevertheless we expect the results of the
$L_{\mathcal{C}}=8$ calculation to be still heavily biased towards the
insulating state. One can regard the critical value $U\approx 0.7\,$eV
for which the gap opens at $L_{\mathcal{C}}=8$ as a lower bound to the
true critical interaction. 

\bibliography{LiMoO}{}

\begin{thebibliography}{113}
\bibitem[{foo({\natexlab{a}})}]{footnote1}
\bibinfo{note}{For a simplification of the calculation, in particular the
  structure of the $k$-mesh, we approximated $\beta$ to $90.0^\circ$. We
  checked that the band structure is indistinguishable along the high-symmetry
  directions.}
\bibitem[{foo({\natexlab{b}})}]{footnote7}
\bibinfo{note}{All densities are given in terms of orbital densities $\langle
  n_i \rangle=\sum\limits_\sigma\langle n_{i\sigma} \rangle$.}
\bibitem[{foo({\natexlab{d}})}]{footnote2}
\bibinfo{note}{Since we start from a spin symmetric calculation
  $H_{\text{Wannier},\alpha\beta}$ is independent of spin $\sigma$.}
\bibitem[{foo({\natexlab{f}})}]{footnote5}
\bibinfo{note}{When comparing results of different techniques (LDA, VCA, DMFT,
  ARPES), we first apply a rigid band shift of $0.03$\,eV to the LDA data to
  account for the $\langle n \rangle \approx 1.44$ filling (which matches the
  as-is ARPES results). In order to be consisten with this LDA data, the VCA
  and DMFT data have to be shifted further by an offset which is determined by
  the way $k$-sums are handled in the respective methods. The chemical
  potential offset of the respective methods, determined for $U=0$\,eV, is
  $\approx0.1$\,eV for VCA and $\approx0.13$\,eV for DMFT.}
\bibitem[{foo({\natexlab{g}})}]{footnote4}
\bibinfo{note}{Note that this may be a minimum, a maximum or in general any
  stationary point in each of the parameters.}
\bibitem[{foo({\natexlab{h}})}]{footnote9}
\bibinfo{note}{Note that for largely de-localized orbitals it is expected that
  the on-site interaction should be slightly larger in a more involved model
  that includes also off-diagonal Coulomb terms.~\cite{Schuler2013}}.
\end{thebibliography}
\end{document}